\documentstyle[preprint,aps]{revtex}
\begin{document}
\draft
\title{Strain effect in silicon-on-insulator materials: \\
Investigation with optical phonons}
\author{J. Camassel, L.A. Falkovsky$^*$ and N. Planes}
\address{Groupe d'Etudes des Semiconducteurs\\
UMR 5650 CNRS-UM2,\\
$cc074 Universit\acute{e}$ Montpellier 2\\
Place E. Bataillon \\
F-34095 Montpellier cedex 5, France}
\date{\today}
\maketitle

\begin{abstract}
We report a detailed experimental and theoretical investigation of the
effect of residual strain, and strain relaxation, which manifests itself at
the Si/SiO$_{2}$ interfaces in commercial silicon-on-insulator (SOI) wafers.
SOI material is made of a single-crystal silicon overlayer (SOL) on top of
an insulator (buried SiO$_{2}$ layer) sitting on a handle silicon wafer.
Infrared reflectivity spectra show that the buried SiO$_{2}$ layer relaxes
continuously when thinning the SOL. At the same time the SOL surface
roughness and the linewidth of optical phonons in Si near the Si/SiO$_2$
interface (probed by micro-Raman specroscopy) increase. In the as-delivered
wafers, this comes from a slight expansion of Si on both sides of the buried
SiO$_{2}$ layer which, conversely, is compressed. Thinning the SOL modifies
these initial equilibrium conditions. To get quantitative results, we have
modeled all our Raman spectra using a theory of inhomogeneous shift and
broadening for optical phonons, which takes into account the phonon
interaction with the static strain fluctuations. From the variation of
linewidth versus interface distance, we have found that the mean squared
strain continues to relax in the bulk of the wafer through a depth on the
order of several $\mu m$. We also show that the SOL surface roughness is
related to strain fluctuations near the Si/SiO$_2$ interfaces.
\end{abstract}


\section{INTRODUCTION}

Silicon-On-Insulator (SOI) technology emerged in the early 80s and became
increasingly popular \cite{MRS},\cite{TG}. Initially limited to space and
military applications where radiation hardness is more important than cost
and performance, SOI material was rapidly considered for PD (Partially
Depleted) and FD (Fully Depleted) C-MOS (Complementary-Metal Oxide
Semiconductor) technology. The net consequence is that IBM announced
recently mass-production of ICs (Integrated Circuits) on SOI using PD-MOS
architectures \cite{Hawaii}.

The main reason for success in the case of ICs on SOI is that unlike
conventional bulk silicon they do not have any direct electrical connection
between the active devices and the underlying substrate. This renders easier
the production of low power / low voltage / high frequency ICs. However,
since FD-MOS remains difficult to master, there is still room for deep
scientific interest. In this work, we report a detailed consideration of the
strain and strain relaxation effects which appear in such non-traditional
bonded silicon substrates.

Whatever the manufacturing technology, all standard SOI wafers have the
appearance shown in Fig. 1. On top is a thin film of single-crystalline
silicon, next a buried oxide film and, finally, a thick (handle) silicon
wafer. Usually the Silicon-OverLayer (SOL) is 0.2 $\mu $m-thick and the
Buried-OXide (BOX) is 0.4 $\mu $m-thick. Basically two different techniques
can be used to produce the buried oxide layer. One is oxygen-ion
implantation, which results in SIMOX (Separation by IMplantation of Oxygen).
The second is hydrophilic bonding of two (previously oxidized) Si wafers,
which results in BESOI (Bond and Etch back SOI) and Unibond \cite{SOITEC}.
Of course, the topmost SOL and the BOX should be defect-free, stress-free,
and uniform in thickness. The handle Si wafer should be also defect-free.

The real situation departs from this ideal viewpoint. From a recent survey
of literature data, the SOL surface roughness ranges from less than 2 $\AA$
in Unibond to 5$\AA$ in high dose SIMOX, whereas the density of dislocations
in the SOL runs from 100 cm$^{-2}$ in Unibond to 10$^{6}$ cm$^{-2}$ in
SIMOX. Since any direct bonding of two oxidized wafers (or any oxygen-ion
implantation followed by a high temperature annealing step) must introduce a
finite amount of stress at the interface \cite{Letavic}, it is obvious that
the primary source of defects in the SOL is the stress located at the Si/SiO$%
_2$ interface. Upon relaxation, this stress leads to the generation of
dislocations which, in turn, affect the silicon parts of the material.

The study of the interaction between dislocations and a strain field is an
old problem. In the recent years, it has been actively pursued (see, for
instance, Refs. \cite{Sch} -\cite{RPM}), but many qualitative observations
are still not satisfactorily understood. For instance, in recent work \cite
{EL}, we have reported some preliminary results concerning the strain in
SOI, which need to be quantitatively discussed. We have used SIMOX and
Unibond materials and, in both cases, started from $\sim $ 200 nm (nominal)
SOL thickness. Then we thinned progressively down to $\sim $ 5 nm and
observed three different phenomena: (i) a change in the 1050 cm$^{-1}$
infrared reflectivity structure associated with the internal vibrational
modes of the SiO$_4$ tetrahedra located in the BOX, (ii) an increase in the
SOL surface roughness probed by AFM (Atomic Force Microscopy) and, (iii) a
variation in the linewidths of the optical phonons in the bulk of the handle
Si-wafer as a function of the distance from the BOX/wafer interface. This
was probed using micro-Raman spectroscopy. It was found that roughly
speaking, given a specific SOL thickness, the linewidth becomes narrower as
the distance from the interface increases. Conversely, given a distance, the
linewidth was larger for the samples with the thinner overlayer. This
behavior corresponds with surface roughness modification. The roughness of
the free surface increases when the thickness of the overlayer decreases.

Since the same distance-to-interface dependence of the linewidth is found in
different kinds of SOI (BESOI and SIMOX materials) it is necessary to
propose a common model to explain the physical origin of this phenomenon.
For instance, because the natural width of a Raman phonon line is mainly
determined by the anharmonic decay of an optical phonon (with frequency $%
\omega _{0}$) into two phonons of opposite wave vectors, additional
contributions near the interface could result from the decay into two
surface phonons or, in the long-wave limit, into two acoustic surface
Rayleigh waves. The excited surface phonons have frequencies $\omega _{0}/2$
and wave vectors $k=\omega _{0}/2s_{t}\zeta $ (where $s_{t}$ is the
transverse-acoustic velocity and the constant $\zeta $, of the order of 0.9,
depends on the elasticity constants of the material). Therefore, in the
direction normal to the surface, the corresponding penetration depth $\delta
\sim 1/k$, being on the order of wavelength, is not larger than a few tens
of lattice parameters $a$. Then the contribution to the Raman spectra must
be of the order of $\delta /d_{s}$, where $d_{s}$ is the diameter of the
laser spot. Such a small contribution is not observable in our experiments.

An alternative reason for broadening could be related to imperfections
localized near the interface. In a single crystal only the phonon modes with
a well defined wave-vector (close to the zone-center because the incident
and scattered photons have very small momenta) can contribute to the Raman
spectra. All departures from this "perfect crystal" conditions are known as
a "relaxation of the momentum selection rule" \cite{TAP}. A simple
(qualitative) description is usually obtained \cite{RWL}, by integrating the
natural (Lorentzian) phonon line shape with a Gaussian distribution function
of phonon states (the width of which is determined by the space scale of the
imperfections). However, such a description does not explain the physical
source of broadening.

In this work we challenge a model which takes quantitatively into account
the scattering of phonons by structural imperfections. The central point is 
strain induced at the Si/SiO2 interface because of the large difference in
thermal expansion coefficients. This results in structural defects (or
misfit dislocations) in the adjacent layers. Away from the interface the
strain relaxes at some finite distance, involving new threading
dislocations. The effect of the smooth part of the strain on the phonon
frequencies can be compared with the influence of pressure. With this
respect, our theory is consistent with the well known results for the
pressure dependence of phonon frequencies. This was previously observed
under homogeneous stress in Si, Ge and GaAS \cite{CBP} or thin films of
cubic SiC on Si \cite{FCP}, \cite{FBC} and results in well-known phonon
shift and splitting.

The specific effect of strain fluctuations is to give rise to a phonon
scattering process which induces additional phonon width and shift. Despite
the fact that the effect of disorder on vibrational modes has been studied
for a long time \cite{KL},  a detailed model has only been proposed
recently. Taking into account the optical phonon scattering due to disorder,
the theory describes the inhomogeneous phonon width associated with phonon
scattering by the strain fluctuations and explains satisfactorily all
relevant experimental results \cite{FBC}.

One important point to notice is that the most distinctive features of the
optical-phonon scattering by static imperfections is the influence of the
DOS singularity at the zone-center. If the optical phonon branch has a
maximum (minimum) value $\omega _{0}$ at the zone-center, the cross section
of the phonon--impurity scattering vanishes at $\omega =\omega _{0}$ and
increases for $\omega ^{2}<\omega _{0}^{2}$ ($\omega ^{2}>\omega _{0}^{2}$,
respectively). Then the Raman line shape as a function of the frequency $%
\omega $ shows asymmetry, being broader on the low (high) frequency side.
The asymmetry is sensitive to the dimensions of the disorder \cite{F}. This
important role of the ''density-of-state'' effect was noticed only recently
for films subjected to laser heating \cite{RHL} and explain satisfactory the
final line shape observed in this work.

Our paper is organized as follows. To analyze the experimental data in
detail, we discuss the theory of probing the strain by Raman spectroscopy in
Sec. II. In Sec. II-A, we recapitulate the equation of motion of optical
phonons in the long-wave approximation and show that the effect of
homogeneous strain gives rise to the splitting and shift of the optical
phonon triplet. In Sec. II-B, we show how the static strain fluctuations
involve the phonon scattering and result in increased phonon linewidths.
Since the increase in linewidth can be larger than the natural phonon width
in the anharmonic decay channels, we consider the effect of fluctuations
self-consistently using Dyson's equation for the phonon Green's function. We
obtain the inhomogeneous broadening and shift in terms of the strain
correlation function. Supposing that the main strain effect is connected to
the dislocations, a two-dimensional form of the correlation function is
used. In Sec. II-C, we discuss the effect of surface roughness. In Sec. III,
we summarize all experimental details and compare them with theory. In
Subsec. III-A we discuss the results obtained from the infrared spectra
collected as a function of SOL thickness for various SOI wafers. We show
that they give evidence of residual permanent strain in the as-delivered
wafers. Within the BOX, the strain relaxes more and more when the SOL
thickness decreases. This is independent of the SOI wafer fabrication
process as well as the thinning method. In Subsec. III-B, we show that this
decrease of residual strain in the BOX correlates directly with an increase
in the Raman linewidth in the handle silicon wafer. This is observed from
micro-Raman spectra collected on the cleaved edges of wafers. Near the
interface Raman spectra show that the threefold degeneracy of the $k\simeq0$
optical phonon is split indicating a large (local) strain. We deduce a
noticeable inhomogeneous contribution to the linewidth, which extends down
to several micrometers. Finally, we show that the surface SOL roughness is
directly related to the strain fluctuations near the BOX/wafer interface.

\section{THEORY}

\subsection{ Summary of strain effect on optical phonons}

In Si, at the $\Gamma $ point of the Brillouin zone the optical phonon
displacements $u_{i}({\bf r},\omega )$ (where $i=x,y,z$) belong to the
threefold representation $\Gamma _{25}^{\prime }$ of the diamond group $%
O_{h}^{7}$. The interaction of these phonons with the static strain $%
\varepsilon _{ij}({\bf r})$ corresponds to the third order terms (which are
linear in $\varepsilon _{ij}({\bf r})$ \ and bilinear in $u_{i}({\bf r}%
,\omega )$ ) in the total energy expression. The first derivatives of the
total energy with respect to $u_{i}({\bf r},\omega )$ give then the equation
of motion 
\begin{equation}
\left( (\omega _{0}^{2}+s^{2}\Delta -i\omega \Gamma ^{(int)}-\omega
^{2})\delta _{ij}+V_{ij}({\bf r})\right) u_{j}({\bf r},\omega )=0
\label{oeq}
\end{equation}
where the zone-center optical phonon frequency $\omega _{0}=520$ cm$^{-1}$
for Si and the corresponding natural width $\Gamma ^{(int)}$ is about 3 cm$%
^{-1}$. The optical branches in cubic crystals are dispersive and, for
instance, the dispersion of the $x-$polarized optical branch has the form: $%
\omega ^{2}=\omega
_{0}^{2}-s_{1}^{2}k_{x}^{2}-s_{2}^{2}(k_{y}^{2}+k_{z}^{2}) $. The neutron
scattering experiments give an estimate of the dispersion parameters $%
s_{1}\approx s_{2}\approx 0.8\cdot 10^{5}$ cm/s for Si. This shows that the
dispersion term can be written in a more simple (isotropic) form $%
s^{2}\partial ^{2}/\partial x_{i}^{2}$ in the vicinity of the branch maximum
at the Brillouin-zone center.

The interaction $V_{ij}({\bf r})$ describes the changes in "spring
constants" due to strain $\varepsilon_{ij}({\bf r})$: 
\begin{equation}  \label{str}
V_{xx}({\bf r})=\lambda_{xxxx}\varepsilon_{xx}({\bf r})+
\lambda_{xxyy}[\varepsilon_{yy}({\bf r})+ \varepsilon_{zz}({\bf r})] \quad 
\text{and} \quad V_{xy}({\bf r})=\lambda_{xyxy}\varepsilon_{xy}({\bf r}),
\end{equation}
where we take into account the fact that only three independent components $%
\lambda_{ijkl}$ exist in a cubic crystal. The other elements of $V_{ij}({\bf %
r})$ can be obtained by circular permutations of the indices in Eq. (\ref
{str}). It is evident that they are of order $\omega_0^2$. From experiments
with homogeneous stress \cite{CBP} we find the coefficient values $%
\lambda_{xxxx}\equiv p=-1.6~\omega_0^2$, $\lambda_{xxyy}\equiv q=
-2.2~\omega_0^2 $ and $\lambda_{xyxy}\equiv 2r=-1.3~\omega_0^2$ , where the
generally accepted notation $p, q, r$ is used.

Because of the strain fluctuations, all quantities have to be averaged over
physically small distances (which are still large on a microscopic scale).
The averaged strain $\langle \varepsilon_{ij}\rangle$ conserves cubic
symmetry in the plane parallel to the interface, but depends smoothly on the
distance to the interface. In Fig. 1 this is chosen as the $z-$direction. We
do not write this dependence in explicit form. The average components are
such that: $\langle \varepsilon_{xy}\rangle =\langle \varepsilon_{xz}\rangle
=\langle \varepsilon_{yz}\rangle =0$ and $\langle \varepsilon_{xx}\rangle
=\langle\varepsilon_{yy}\rangle $, which gives for the shifted phonon
frequencies at the Brillouin-zone center 
\begin{equation}  \label{perfr}
\omega _z^2(k=0)=\omega _0^2+\lambda _{xxxx}\langle \varepsilon_{zz}\rangle
+2\lambda _{xxyy}\langle \varepsilon_{xx}\rangle
\end{equation}
for the phonon polarization normal to the interface and 
\begin{equation}  \label{parfr}
\omega _x^2(k=0)=\omega _y^2(k=0)=\omega _0^2+(\lambda _{xxxx}+\lambda
_{xxyy})\langle \varepsilon_{xx}\rangle + \lambda _{xxyy}\langle
\varepsilon_{zz}\rangle
\end{equation}
for the phonons polarized in the plane of the interface. This is nothing but
the well-known phonon spliting into a $z-$singlet and an $xy-$doublet under
the effect of a homogeneous strain \cite{CBP} - \cite{FBC}.

\subsection{Effect of strain fluctuations on the optical-phonon linewidth}

The contribution to the linewidth arises in the second-order perturbation
theory for the strain components. It could be calculated using "the golden
rule" of quantum mechanics. However, due to the singularity in the phonon
density-of-states at $\omega=\omega_0$, this contribution is strongly
frequency-dependent and may be larger than the natural phonon linewidth.
Therefore we calculate it self-consistently using a Green's function
technique \cite{AGD}. For the first-order Raman scattering, the
corresponding cross section is bilinear in the phonon displacements. Then
the Raman cross section 
\begin{equation}  \label{crs}
\frac{d\sigma}{d\omega^{(s)} d\Omega^{(s)}}\propto \frac{ e^{(i)}_{\beta
}e^{(s)}_{\gamma } e^{(i)}_{\beta^{\prime}}e^{(s)}_{\gamma ^{\prime}}
g_{\alpha \beta \gamma}g_{\alpha ^{\prime}\beta ^{\prime}\gamma ^{\prime}} 
} { 1 - \exp(- \hbar\omega /k_BT)}~{\rm Im}\, D_{\alpha \alpha ^{\prime}}(%
{\bf k },\omega),
\end{equation}
is determined by the imaginary part of the phonon Green's function $%
D_{\alpha \alpha ^{\prime}}({\bf k },\omega)$, where $\omega$ and ${\bf k }$
have the sense of frequency and momentum transfers, $e^{(i)}_{\beta }$ and $%
e^{(s)}_{\gamma }$ are the polarization vectors of the incident and
scattered photons, respectively, and $g_{\alpha \beta \gamma}$ is the $\beta
\gamma-$component of the Raman polarizability tensor for the $\alpha-$phonon
branch. It is easy to see that, for the $\Gamma_{25}^{\prime}$
representation, the tensor $g_{\alpha \beta \gamma}$ is completely
antisymmetric and reduces to the component $g_{x y z}$. This means that,
when the incident and scattered fields are polarized along the $x$ and $y$
axes, respectively, only $z-$polarized phonons can be excited.

The phonon Green's function has to be averaged over the strain distribution.
The calculation has been performed in details in a previous work\cite{FBC}
and we shall simply summarize the results. The average phonon Green's
function has the diagonal form: 
\begin{equation}
D_{jj}({\bf k},\omega )=\left( \Omega _{j}^{2}({\bf k},\omega
)-s^{2}k^{2}-i\omega \Gamma _{j}({\bf k},\omega )-\omega ^{2}\right) ^{-1},
\label{df}
\end{equation}
where the unknown real functions $\Omega _{j}({\bf k},\omega )$ and $\Gamma
_{j}({\bf k},\omega )$ obey the system of Dyson's equations 
\begin{equation}  \label{br}
\Omega _{j}^{2}({\bf k},\omega )-\omega _{0}^{2}-\langle V_{jj}\rangle
-i\omega (\Gamma _{j}({\bf k},\omega )-\Gamma ^{(int)}) = -\sum_{q}\frac{%
W_{jm}({\bf q}-{\bf k})}{\Omega _{m}^{2}({\bf q},\omega )-s^{2}q^{2}-i\omega
\Gamma _{m}({\bf q},\omega )-\omega ^{2}}.
\end{equation}
The average interaction $\langle V_{jj}\rangle $ is defined in Eqs. (\ref
{str}) - (\ref{parfr}) \ and the correlation function in real space is 
\begin{equation}
W_{jm}({\bf r-r^{\prime }})=\langle \delta V_{jm}({\bf r})\delta V_{mj}({\bf %
r^{\prime }})\rangle  \label{korf}
\end{equation}
with $\delta V_{jm}({\bf r})=V_{jm}({\bf r})-\langle V_{jm}\rangle .$

Several points should be emphasized. First, one can see from Eqs. (\ref{crs}%
) and (\ref{df}) that the line center as a function of the frequency
transfer $\omega $ is determined by the equation $\Omega _{j}^{2}({\bf k}%
,\omega )-s^{2}k^{2}=\omega ^{2}$, where ${\bf k}$ is the momentum transfer.
Here we can put $k=0$, since the photon wave vector is small. Second, in the
limiting case $\Gamma _{j}({\bf k},\omega )\rightarrow 0$, the bypass around
the pole in the right-hand side of Eq. (\ref{br}) gives the Dirac
delta-function $\delta (\Omega _{m}^{2}({\bf q},\omega )-s^{2}q^{2}-\omega
^{2})$. We then arrive at the well-known golden rule and see that the strain
fluctuations make a contribution into the linewidth only for $\omega
^{2}<\Omega _{m}^{2}$, i. e. on the low-frequency side of the Raman line.
The opposite situation occurs when the phonon branch has a minimum at the
center of the Brillouin zone\cite{F} but, in both cases, the line shape
shows asymmetry.

\subsection{Simplest model of two-dimensional disorder}

Equations (\ref{br}) can be simplified if we assume that the correlator $%
W_{jm}({\bf q})$ has a constant value $W_{jm}({\bf q}=0 )$ in the region $%
q<1/r_0$ and vanishes elsewhere. On the other hand, the strain fluctuations
may be considered as imperfections where $W_{jm}({\bf q}=0 )$ is
proportional to the concentration of imperfections and $r_0$ is the domain
size in which a phonon interacts with the imperfection. The final form of
Eqs. (\ref{br}) depends on the dimensionality of the imperfections. In our
previous work \cite{FBC} we have concluded that the imperfections involved
by interfaces are mainly two-dimensional, i. e. that they behave like
dislocations. In that case, $r_0$ is the dislocation core and the
correlation function $W({\bf r})$ in Eq. (\ref{korf}) depends on the
two-dimensional vector ${\bf r}_{\bot}$. Then Eqs. (\ref{br}) have the form: 
\begin{eqnarray}  \label{wu}
\Omega _{j}^{2}(\omega)- \omega _0^{2} - \langle V_{jj}\rangle-i\omega
(\Gamma _{j}(\omega)-\Gamma^{(int)} )  \nonumber \\
= \omega _0 B_{jm} \left(\frac{1}{2} \log{\ \frac{(s^2/r_0^2+\omega^2-
\Omega^2_m(\omega))^2+ \omega^2\Gamma^2_m(\omega)} {(\omega^2-\Omega^2_m(
\omega))^2+\omega^2\Gamma^2_m(\omega)}}\right. \\
\left.- i \arctan{\frac{\omega^2-\Omega^2_m(\omega)+s^2/r_0^2} {
\omega\Gamma_m(\omega)}} +i\arctan{\frac{\omega^2-\Omega_m^2(\omega)} {
\omega\Gamma_m(\omega)}}\right),  \nonumber
\end{eqnarray}
where $B_{jm}=W_{jm}({\bf k}_{\bot}=0)/4\pi s^2\omega _0$, and the sum can
be taken over two split modes, which are the normally polarized singlet and
the doublet parallel to the interface. Notice that the functions $\Omega
_{j}(\omega)$ and $\Gamma _{j}(\omega)$ are independent of ${\bf k}$ (we
omit ${\bf k}$ in the arguments) because of its smallness in comparison to
the value $\sqrt{\omega_0 \Gamma}/s \simeq (\pi/a)\sqrt{\Gamma/\omega_0}$
which is essential in the integral (\ref{br}). Again $a$ is the lattice
parameter.

Let us estimate $B_{jm}$. According to Eqs. (\ref{oeq}) and (\ref{str}), $%
V(r=0)=\lambda\varepsilon\sim\omega_0^2 \varepsilon$, where we omit the
tensor indices. We have for the Fourier component (\ref{korf}) $%
W_{jm}(k=0)\sim 2\pi\omega_0^4 (\delta\varepsilon)^2r_0^2$ and for the
phonon-strain scattering probability 
\begin{equation}  \label{est}
B\sim \omega_0^3(r_0\delta\varepsilon /s)^2
\end{equation}
where $\delta\varepsilon$ is the strain fluctuation.

If we assume that the strain fluctuations are mainly induced by
dislocations, then 
\begin{equation}
B=cv^{2}(k=0)/4\pi \omega _{0}s^{2}.  \label{este}
\end{equation}
In this expression, $c$ is the dislocation concentrations, $%
v(k=0)=gr_{0}^{2}\omega _{0}^{2}$ is the phonon-dislocations interaction
where $g$ is a dimensionless constant of the order of unity and $r_{0}$ is
the radius of a dislocation core.

In the following we will solve numerically Eqs.(\ref{wu}) in $%
\Omega_{j}(\omega )$ and $\Gamma _{j}(\omega )$ to find the values of $B_{jm}
$ and $r_{0}$ which give a fit of Eqs. (\ref{crs}) and (\ref{df}) to the
experimental Raman spectra. Notice that the effect of the strain
fluctuations on the line shape is qualitatively evident from Eqs.(\ref{wu}).
For instance, if there is only one phonon mode, its width $\Gamma (\omega )$
is governed by the equation 
\begin{equation}
\Gamma (\omega )-\Gamma ^{(int)}=\frac{\omega _{0}B}{\omega }\left( \arctan {%
\ \frac{\omega ^{2}-\Omega ^{2}(\omega )+s^{2}/r_{0}^{2}}{\omega \Gamma
(\omega )}}-\arctan {\frac{\omega ^{2}-\Omega ^{2}(\omega )}{\omega \Gamma
(\omega )}}\right) .  \label{wu1}
\end{equation}
Let us consider the case of small $r_{0}$ ($r_{0}<s/\sqrt{\omega \Gamma }$).
The terms in the right-hand side of Eq. (\ref{wu1}) compensate for each
other on the high-frequency side ($\omega ^{2}>\Omega ^{2}$) of the Raman
line, where the line has a nearly Lorentzian form with a width $\Gamma
\simeq \Gamma^{(int)}$. For the low-frequency side ($\omega ^{2}<\Omega ^{2}$%
), these two terms give a contribution to the linewidth which is
proportional to the squared strain fluctuations. For the situation of a
minimum in the phonon branch at the center of the Brillouin zone, the Raman
line drops more rapidly on the low-frequency side. This effect has origin in
the singularity of the phonon density-of-states at the branch extremum. In
the opposite case, $r_{0}\gg s/\sqrt{\omega \Gamma }$, the lineshape is
symmetric but non-Lorentzian.

Since $\Gamma (\omega )$ is $\omega -$dependent, we must connect its value
with the full width at half maximum (FWHM) of the experimental spectra. The
simplest way is to use $\Gamma (\omega )$ at the line peak, where the
equation $\Omega (\omega )=\omega $ is satisfied. In the case where $\Gamma $
is independent of $\omega $, this definition gives the FWHM. The first term
in the parenthesis Eq.(\ref{wu1}) remains only at the line peak. Finally,
because the difference between the line positions $\Omega (\omega )$ and $%
\omega _{0}$ (including and ignoring the effect of disorder, respectively)
is small, we obtain for $\Gamma $ at the line peak 
\begin{equation}
\Gamma -\Gamma ^{(int)}=B\arctan \frac{s^{2}}{r_{0}^{2}\omega _{0}\Gamma }.
\label{wu2}
\end{equation}
Notice that, in the limiting case $r_{0}\ll s/\sqrt{\omega \Gamma },$ Eq.(%
\ref{wu2}) coincides with the result of the perturbation theory $\Gamma
=\Gamma ^{(int)}+\pi B/2$. In the opposite case $\arctan
(s^{2}/r_{0}^{2}\omega _{0}\Gamma )=s^{2}/r_{0}^{2}\omega _{0}\Gamma $ and
we arrive at a quadratic equation which gives 
\begin{equation}
\Gamma =\frac{\Gamma ^{(int)}}{2}+\sqrt{\left( \frac{\Gamma ^{(int)}}{2}%
\right) ^{2}+\frac{Bs^{2}}{\omega _{0}r_{0}^{2}}}.  \label{wu3}
\end{equation}
This shows that, in the general case, the effect of disorder which is
represented by the parameter B, is not simply additive with the natural
width $\Gamma ^{(int)}$.

\subsection{Surface roughness induced by dislocations}

Because of the interface strain and related dislocations, there must be a
finite surface roughness contribution. Due to the very complicated nature of
the interactions between the surface and dislocations it seems impossible to
reach a high degree of sophistications in this problem. However, we can
estimate the mean roughness using a qualitative approach. We suppose that
the strain has large fluctuating components and calculate the mean square
roughness.

Any rough plane may be conceived as a set $\zeta ({\bf s})=\sum_{n}b({\bf s}-%
{\bf s}_{n})$ of randomly located ''hills'' at points ${\bf s}_{n}$, where $%
{\bf s}$ is a two-dimensional vector of the free surface plane. If we
suppose that the hills originate from dislocations, the height $b$ must be
on the order of the Burgers vector $b_{0}$. We specify $l$ as the size of
the hills (i.e. the domain size where the function $b({\bf s}-{\bf s}_{n})$
has a nonzero value) and, by averaging over the distribution of points ${\bf %
s}_{n}$, we obtain the roughness correlation function 
\[
\zeta_2({\bf s}-{\bf s}^{\prime})=<\zeta ({\bf s})\zeta ({\bf s}^{\prime})>
=c\sum_q \mid b({\bf q})\mid^2 e^{i{\bf q}({\bf s}-{\bf s}^{\prime})}, 
\]
where $b({\bf q})\sim b_0l^2$ is the Fourier component of the function $b(%
{\bf s})$ and $c$ is the hill concentration at the surface.

The mean square roughness is given by $\zeta_2({\bf s}={\bf s}%
^{\prime})\simeq c b_0^2l^2$ which is connected to the strain relaxation by
the dislocation concentration $c$. Of course some initial roughness $%
\zeta_0^2$ does exist, which is not strain-dependant. Taking both
contributions into account, we write for the effect of strain on the surface
roughness 
\[
\zeta^2=\zeta_0^2+c b_0^2l^2. 
\]
This expression and Eqs. (\ref{este}), (\ref{wu2}) enable us to compare the
change in roughness with the phonon linewidth. We get 
\begin{equation}  \label{rou}
\Gamma -\Gamma ^{(int)}=A(\zeta^2-\zeta_0^2) \arctan \frac{s^{2}}{%
r_{0}^{2}\omega _0\Gamma },
\end{equation}
where $A=g^2r_0^4\omega_0^3/4\pi s^2b_0^2l^2.$

\section{Experimental results and comparison with theory}

Most experimental details have already been given in \cite{EL} and will not
be repeated here. To summarize briefly, three different (commercial) 4-inch
wafers from SOITEC\cite{SOITEC} were considered. One wafer was high dose
SIMOX and  two were Unibond. In all three cases, the nominal SOL thickness
was 0.2 $\mu$m. In two cases (one Unibond wafer and the SIMOX wafer), a wet
SO (Sacrificial Oxidation) process was used. This resulted in two series of
samples denoted A and B, respectively. In order to evaluate the effect of
the repetitive high temperature oxidation steps (at 1050 C) on the final
results, a low temperature oxygen-assisted IBE (Ion Beam Etching) process
was applied to the second Unibond wafer. This resulted in the third series
of samples (samples C).

\subsection{ Influence of the SOL thickness on the IR reflectivity spectra}

To control the SOL thickness\cite{CRS} infrared reflectivity spectra were
systematically collected at room temperature with a Brucker IFS 66v Fourier
Transform Spectrometer fitted with a DTGS detector. Measurements were made
in the spectral range 400 to 7500 cm$^{-1}$ and a standard "near normal
incidence" configuration (finite value of the angle of incidence close to $%
20^0$) was used. The direction of incidence was in the $yz-$plane of Fig. 1
and no attempt was made to polarize the incident light.

Apart from the change in interference patterns which gives the final sample
thicknesses, a strong effect was found in the 1000 - 1100 cm$^{-1}$ range.
This is the narrowing of the SiO$_{2}$-related feature displayed in Fig. 2
for samples A. It has nothing to do with the change in the interference
pattern, but correlates with the change in SOL thickness. Notice that there
is a large effect when reducing the SOL from 203 to 95 and 42 nm. Notice
also that, starting from 42 nm, it seems to saturate.

The 1070 - 1080 cm$^{-1}$ doublet structure is well known. It manifests
itself in every silicate-like material (not only in crystals, \cite{SP}-\cite
{CGP}, but also in vitreous silica \cite{KIR} or, even, in sol-gel derived
glasses \cite{KPK}) and corresponds with internal vibrations of the building
SiO$_4$ tetrahedra. It is very important to notice the stability of the
corresponding frequency which does not change in going from $\beta-$ to $%
\alpha-$quartz \cite{JWM} or, even, when Si is replaced by Al or P in AlPO$%
_4 $, which is the ternary analog of quartz \cite{CGP}. In the literature it
has been classified as an AS (Asymmetric Stretching) internal mode \cite{KIR}%
, \cite{KPK} in which one oxygen atom bonded to two neighboring Si moves
parallel to the Si-Si direction. This oxygen motion is strongly infrared
active and, despite the lack of long-range ordering, in many cases the
associated LO frequency resolves around 1260 cm$^{-1} $.  Due to the finite
value of the angle of incidence, this feature appears also in the present
work.

Concerning Fig. 2, a second point should be emphasized. In the specific case
of a thin SiO$_2$ film on Si, one expects two equivalent in-plane $x-$ and $%
y-$modes and only one out-of-plane $z-$mode. In other words, because we are
dealing with a thin layer in which the threefold degeneracy is lifted, one
expects to find a doublet structure with a 1 to 2 intensity ratio. This is
exactly what is found. Starting from one broad (unresolved) feature for
nominal SOL thickness 200 nm, one goes rapidly to two distinct vibrational
modes when the SOL thickness reduces below 40 nm. From the corresponding
intensity ratio, we deduce that the $z-$mode appears lower. Since the
average energy difference with respect to the $x,y-$modes comes from the
thermal strain, the strain is compressive. Making the length of the Si-O
bonds oriented in the $xy-$plane shorter, it shifts the $x,y-$modes to
higher energy. Of course there are some finite deviations with respect to
the average energies and, since we find evidence of a clear re-ordering
(within the thin dielectric film) when thinning the SOL, we must conclude
that the SOI system is not at all rigid but, instead, at the atomic scale
constitutes a rather flexible medium.

Since the simplest way to shift the frequency of an internal mode is to
change the bond length, our results suggest that we are dealing with oxide
relaxation. To explain from the very beginning the narrowing of the Si-O
related feature, one must assume an initial (thermal strain induced)
distortion which relaxes when the SOL thickness decreases. Starting from a
broad feature ($\sim $100 cm$^{-1}$) for the as-delivered material (broad
distortion of the equilibrium bond length) the narrowing of the Si-O bond
signature demonstrates a reordering of the Si-O distances as the SOL
thickness decreases. This is true from the very beginning, when thinning the
SOL from 200 to 100 nm, and does not depend on the type of material
investigated. This observation has been checked on SIMOX (series B) and
similar results were found.

Finally, to check that this was not due to any parasitic effects of the high
temperature SO steps, we repeated the thinning process using oxygen-assisted
IBE (series C). Again, similar results were found. This lends support to the
proposition that we are dealing with real and intrinsic properties: the
experimental improvement found in the oxide relies on an internal stress
relaxation mechanism, where the stress originates only from the difference
in thermal expansion coefficients between Si and SiO$_2$ ($\sim$80\% at room
temperature). There is no artifact associated with the high temperature SO
steps and, to the best of our knowledge, this provides the first clear
experimental evidence that all SOI materials are plastic systems, soft
enough to relax when the SOL thickness decreases.

To estimate the relaxation we write the interatomic potential energy
proportionally to $r^{-n}$, where $n$ should be of the order of 5 for a
typical oxide. Because the relative change of the vibration frequency is $%
\delta\omega/\omega = (n+2)\delta r/2r$ and the linewidth of the
reflectivity signal changes by around 50 cm$^{-1}$ when thinning the SOL, we
deduce that the average change in Si-O bond length is a reduction of about $%
1\%$.

\subsection{Raman investigation of nominal and thinned SOI}

We have mainly shown that, because of internal stress relaxation, the BOX
properties improve when thinning the SOL. Now we consider what happens in
the silicon overlayer and wafer. First we investigated in details the Si
handle-wafer by using micro-Raman spectroscopy. We use a Jobin-Yvon ISA T
64000 spectrometer fitted with an Olympus microscope and a cooled CCD
detector. The 488 nm line of a Spectra-Physics argon-ion laser was used as
the excitation frequency. To achieve in-depth resolution, a transverse
backscattering configuration was used \cite{FBC} with both the incident and
scattered light propagating along the $x-$direction of Fig.1. Because of the
large experimental aperture we probe, both, $y$ and $z$ polarized phonons.

Typical results are shown in Figs. 3 to 6 for the series of samples A
(Unibond). Fig. 3 corresponds to nominal material with SOL thickness 203 nm
(standard value), Fig. 4 with 42 nm (thinned), Fig. 5 with 11 nm (thinned)
and, finally, Fig. 6 with 5 nm. Given a specific SOL thickness, the
different spectra correspond with the different laser spot distance from the
wafer surface. To quantify in full details the different contributions, they
have been fitted using the theory of phonon interaction with static strain
fluctuations presented in Sec. II. The results are shown as solid lines on
the different figures. The complete series of parameters is given in Table 1.

The fits with Eq. (\ref{crs}) and Eq. (\ref{df}) were made simultaneously
for all spectra. The best value of the parameter $r_{0}\omega _{0}/s=16$
gives a reasonable value for the correlation radius (or the dislocation
core) $r_{0}=1.3\AA $. Eqs. (\ref{wu}) were solved numerically to obtain the
matrix elements $B_{ij}$, which determine the inhomogeneous linewidth and
the shift of the lines. One example of the solution (functions $\Gamma
_{j}(\omega )$ and $\Omega _{j}(\omega )-\omega _{0}-\langle V_{jj}\rangle /2
$) is presented in the right panel of Fig. 4 for the $z-$phonon ($j=z$). The
resulting values (i.e. the matrix elements $B_{xx}$ and $B_{zz}$) are given
in Table I for samples A.

Every time, focusing far from the interface (a typically value is around 5 $%
\mu $m for the complete series of samples) one finds an identical (standard)
''bulk'' Raman linewidth for the silicon handle wafers. In the following
this is taken as ''reference material'' and the line position for the
reference material (520 cm$^{-1}$) is shown by the vertical dash-dotted line
as a guide for the eye. Moving toward the surface (first to 2 $\mu $m and,
then, to 1 $\mu $m) has two consequences: i) one shifts significantly the
phonon frequency to lower energy and ii) one broadens considerably the width
of the Raman peak, especially in the cases of the thinner SOL. This is
evident starting from 42 nm.

A first contribution to the phonon lineshift towards the low frequencies
comes as a result of the expansion of the Si lattice near the Si/SiO$_2$
interface \cite{CBP}, \cite{FBC}. This homogeneous shift is described by
Eqs. (\ref{perfr}) - (\ref{parfr}) but, because of the small energy
difference, the strain-induced splitting into a singlet and a doublet states
resolves only for the thinnest SOL thicknesses (5 and 11 nm, respectively:
see Table 1).

The second contribution comes from the strain relaxation mechanism which
manifests itself experimentally when probing at different positions along
the $z-$direction. It is frequency dependent and gives the inhomogeneous
linewidth and shift contributions displayed in the right panel of Fig. 4.
Notice that, when focussing on the active part of the SOI wafers (upper part
of the handle Si wafer) the inhomogeneous linewidth and shift contributions
at the line center become very large. They range from 1.1 cm$^{-1}$ (width)
and +0.14 cm$^{-1}$ (shift) for the 42 nm SOL sample to 6.1 cm$^{-1}$ and
+0.36 cm$^{-1}$, respectively, for the 5 nm SOL sample (see again Table I).
The inhomogeneous contribution to the line-shift remains however smaller
than the homogeneous ones: --0.9 cm$^{-1}$ for the 42 nm SOL sample and
--1.2 cm$^{-1}$ for the 5 nm SOL sample.

Similar results are found for SIMOX. They are displayed in Fig. 7 for the
SOL thickness 50 nm. The natural width is slightly larger (3.4 instead of
3.1 cm$^{-1}$ for Unibox) but the correlation radius has the same value $%
r_{0}\omega _{0}/s=16$. The main distinction is that the depth of strain
relaxation is larger: there is still a noticeable difference in the
linewidths at distances 5 and 20 nm to the interface (see Table II).
Finally, some homogeneity at the interface is found. Two spectra (last rows
of Table II) collected from different points at the interface show
practically the same characteristics. However other regions were detected,
where more dramatic differences do exist. We shall come back to this point
in Sec. IV (Conclusions).

To summarize this section, we have found from our analysis of the linewidth
variation versus distance that all strain-induced effects (like the
homogeneous and inhomogeneous shifts and broadening) have a maximum near the
Si/SiO$_2$ interfaces. Because they depend directly on the SOL thickness,
they should affect primarily the thinned SOLs. The surprising effect is that
they affect also the upper part of the handle wafers and relax through a
depth on the order of several $\mu $m.

\subsection{Surface roughness and Raman width}

From the Raman investigations, we have found that the morphology of the
upper part of every (nominal and thinned) SOI wafer experiences a complex
stress distribution. Moreover, the strain in the wafer near the BOX/wafer
interface becomes larger when thinning the SOL. In this section we show that
independent evidence can be found from the consideration of root mean square
roughness of the SOL surface. We reported already that there is a strong
degradation of the surface when the SOL thickness decreases \cite{EL}.
Typically, decreasing the SOL thickness from 203 to 5 nm, one increases the
mean square AFM (Atomic Force Microscopy) roughness from 1.5 to 4 \AA\ RMS.

To show that the Raman width and mean surface roughness have the same origin
(inhomogeneous strain and dislocations) we show in Fig. 8 a correlation of
the Raman linewidths (cm$^{-1}$) obtained at the surface of wafers A with
different SOL thickness with AFM data ($\AA $). The solid lines are plots of
Eq. (\ref{rou}). The experimental values are shown by dots. A single
parameter $A=2.1$ cm$^{-1}\AA ^{-2}$ was chosen to fit the values of the $z-$%
phonon width and roughness in the case of the SOL thickness of 5 nm. Notice
that the doublet splitting resolves only for the 5 and 11 nm SOL thickness.
The value $A=0.28$ cm$^{-1}\AA ^{-2}$ for the $x-$phonon was obtained in the
comparison of the interaction constants $B_{xx}$ and $B_{zz}$ from Table I.
From these data, it is interesting to estimate the parameter $l $ of the
surface roughness using this value of A ($r_{0}$ is obtained in this work
and $s$ is known). Taking into account the expression of $A$ given after (%
\ref{rou}) and putting $g=1$ and $b_{0}=1\AA $, we get a very reasonable
value $l=88\AA $ for the average diameter of the hills. The main difference
from one thinning process to the next one is in the initial roughness. With
this respect, all results obtained using SO (1.2 \AA, series A) appear
better than the one obtained with IBE (2 \AA, series C).

\section{ Conclusions}

It has been known for a very long time that any direct bonding of two
oxidized wafers (or any oxygen-ion implantation followed by a high
temperature annealing step, for instance) introduces a finite stress at the
Si/SiO$_2$ interface. This comes directly from the difference in thermal
expansion coefficients between Si and SiO$_2$ ($2.6\cdot 10^{-6}$ K$^{-1}$
and $0.56\cdot 10^{-6}$ K$^{-1}$, respectively \cite{TG}). Because this
constitutes a hard point of the technology, low temperature bonding using
low viscosity oxides like BPSG (BoroPhosphoroSilicate Glass) or SOG (Spin on
Glass) are under active development. In some cases they will offer
alternative solutions. The problem of process-induced defects is totally
different. To the best of our knowledge, it is much less documented. For
instance there is no clear report of a critical SOL thickness, similar to
the critical layer thickness encountered in hetero-epitaxy. Above such a
critical thickness, the SOL material would be homogeneously strained but
stable. It would strongly relax around that thickness, with a maximum in
inhomogeneous behavior. Finally, it would be heavily strained, but more
homogeneous and again stable below that value. Our experimental results
suggest that such a critical thickness does exist. It is about 40 to 50 nm
and corresponds with the experimental situations illustrated in Fig. 2
(BOX), Fig. 9 (Unibond) or Fig. 10 (SIMOX). In all three cases we find a
critical regime where, simultaneously, the BOX achieves a better relaxation
while there is a very large inhomogeneity in the SOL. The spectra displayed
in Figs. 9 and 10 (solid lines are again theoretical curves) could be
understood in the terms of phonon splitting of the order of 7 to 15 cm$^{-1}$
due to a very high local strain in some finite parts (or domains) at the
interface.

To summarize, we have evidenced in this work that SOI is not a perfectly
strain-relaxed system. It behaves more like a balanced-strain structure
schematically drawn in Fig. 11. The interesting point is that: i)
compressive strain (about 1 \% in absolute value) is always present in the 
buried oxide of as-delivered SOI wafers. This strain diminishes  when the SOL
thickness decreases; ii) on contrary, tensile strain exists in the silicon
overlayer. Dislocations near the interface release the strains and manifest
themselves in final surface roughness (up to 4 \AA\ RMS). The dislocations
can be considered as the origin of the strain fluctuations which result in
optical phonon scattering; finally, iii) tensile strain and strain
fluctuations (again on the order of 10$^{-2}$) also exist in the bulk of the
underlying wafer. Every time, the strain originates from the Si/SiO$_2$
interfaces and comes from the difference in bond length and thermal
expansion coefficient between the two materials. Since both the Si-overlayer
and the bulk Si wafer keep the buried oxide stressed, when thinning the SOL
the strain decreases in the buried oxide and increases in the handle wafer.
The overall strain relaxes very slowly and extends from the BOX/wafer
interface over a few micrometers range. This is in contradiction to the most
common belief that the SOL/BOX/underlying handle wafers are made of very
stable and strain-free materials. As a matter of fact, SOI wafers experience
more equilibrium conditions which depend on every stage of the technology
road.

{\bf ACKNOWLEDGMENTS }

This work was supported in part by the EU commission under contract BRPR
CT96 0261. We greatly thank Martin Eickoff (Daimler-Chrysler, Munich) and Y.
Monteil (Universit\'{e} de Lyon) for sending us the different samples
investigated in this work. We also thank Y. Stoemenos from Aristotle
University of Thessaloniki (Greece) as well as B. Aspar and H. Moriceau from
LETI-CEA (France) for expert discussions about SOI related problems.
Finally, one of us (L.F.) thanks the French Ministry of Education for
partial support during the course of this work.

* permanent address: Landau Institute for Theoretical Physics, Russian
Academy of Sciences, Kosygina 2, Moscow 117 334, Russia. 

\newpage {\bf Figure captions}: \\[0.3cm]Fig. 1. Schematic drawing of the
Raman back-scattering geometry used in this work to investigate the strain
relaxation in the silicon-on-insulator (SOI) system. The strain relaxation
is probed by displacing the laser spot in the $z$-direction. \newline
Fig. 2. Infrared reflectivity spectra collected on SOI (Unibond) material
for different SOL thickness (see labels on the curves).\newline
Figs. 3. Experimental Raman spectra collected on Si-wafers with SOL
thickness 203 nm. The SOL is separated by a 400 nm oxide film from the
handle wafer. The distance from the interface (5, 2, 1, and 0 $\mu $m ) of
the laser spot is indicated on the curves. The solid lines are fits to Eqs. (%
\ref{crs}) - (\ref{br}). The results of the fits are given in Table 1. The
natural width is 3.1 cm$^{-1}$ for all lines, the correlation radius $%
r_{0}\omega _{0}/s=16$. \newline
Fig. 4. Same as Fig. 3 for 42 nm Si-overlayer thickness. In right panel, the
inhomogeneous shift and linewidth are shown as a function of the frequency.
The position of the line peak in fully relaxed material is shown as
dash-dotted line. The inhomogeneous shift (right panel) is positive, whereas
the total shift (left panel) resulting from the strain is negative. 
\newline
Fig. 5. Same as Fig. 3 for 11 nm SOL thickness. \newline
Fig. 6. Same as Fig. 3 for 5 nm SOL thickness. \newline
Fig. 7. Same as Figs. 3-5 but now for a SIMOX wafer with 50 nm Si-overlayer
thickness. Results of the fits are given in Table II. The natural width is
3.4 cm$^{-1}$ for all lines, the correlation radius $r_{0}\omega _{0}/s=16$. 
\newline
Fig. 8. Linewidths of the two components of the optical phonon triplet in Si
near the Si/SiO$_{2}$ interface as a function of the root mean square (RMS)
roughness of the SOL surface. The lines are the result of the theoretical
model (see Eq. (\ref{rou})). The points give the width found from the fit of
spectra to the theory. The corresponding RMS value is extracted from AFM
measurements for the various thickness (203, 95, 42, 11, and 5 nm) of the
Si-overlayer. \newline
Fig. 9. Experimental Raman spectra (for Unibond wafers A) collected for
different points at the BOX/wafer interface (upper line: normal situation;
bottom lines: specific points discussed in the text). The thickness of the
Si-overlayer is 42 nm. \newline
Fig. 10. Full series of Raman spectra (for SIMOX) collected up to 5 $\mu $m
from a specific point at the BOX/wafers interface. The thickness of the
Si-overlayer is 50 nm. \newline
Fig. 11. Schematic drawing of the in-plane strain in the Si-overlayer, the
buried oxide and the handle silicon wafer for two different SOL thickness: 
nominal value 200 nm (solid line) and thinned SOL (dashed line). \newline
\begin{table}[tbp]
\caption{Zone-center optical phonon frequencies and corresponding linewidths
(all values are in cm$^{-1}$) in the Unibond samples (with various top-layer
thickness) in relation to the distance $z$ from the interface. The results
refer to the fits with the present theory. The adjusted coupling parameters
(the last column) $B_{xx},B_{zz}$ are probabilities for the phonon
intrabranch scattering by strain fluctuations. The splitting of triplet $%
\Gamma _{25}^{\prime }$ into the doublet ($\omega _{x}$) and the singlet ($%
\omega _{z}$) is resolved for samples with 5 nm and 11 nm top-layer
thicknesses.}
\label{tb1}\newpage
\par
\begin{center}
\begin{tabular}{|c|c|c|c|c|}
Sample & Distance from & Lines $\Gamma_{25}^{\prime}$ & Linewidths & 
Interactions \\ 
& interface, $\mu$m & $\omega_{z}; \quad \omega_{x}$ & $\Gamma_{z}; \quad
\Gamma_{x}$ & $B_{zz}; \quad B_{xx}$ \\ \hline
203 nm & 5.0 & 520.0 & 3.10 & 0. \\ 
& 2.0 & 519.76 & 3.21 & 0.20 \\ 
& 1.0 & 519.28 & 3.32 & 0.39 \\ 
& 0 & 519.18 & 3.61 & 0.97 \\ \hline
95 nm & 5.0 & 520.0 & 3.10 & 0. \\ 
& 2.0 & 520.0 & 3.10 & 0. \\ 
& 1.0 & 520.0 & 3.32 & 0.38 \\ 
& 0. & 519.28 & 3.69 & 1.16 \\ \hline
42 nm & 5.0 & 520.0 & 3.10 & 0. \\ 
& 2.0 & 520.0 & 3.21 & 0.19 \\ 
& 1.0 & 519.18; 516.13 & 3.69; 3.26 & 1.16; 0.29 \\ 
& 0. & 519.17; 519.08 & 4.25; 3.42 & 2.51; 0.6 \\ \hline
11 nm & 5.0 & 520.0 & 3.10 & 0. \\ 
& 2.0 & 520.18; 520.15 & 3.32; 3.1 & 0.38; 0. \\ 
& 1.0 & 520.27; 520.22 & 3.86; 3.42 & 1.54; 0.58 \\ 
& 0. & 519.55; 519.13 & 6.23; 3.70 & 9.6; 1.2 \\ \hline
5 nm & 5.0 & 520.0 & 3.10 & 0. \\ 
& 2.0 & 519.73; 519.70 & 3.32; 3.1 & 0.39; 0. \\ 
& 1.0 & 519.26; 519.27 & 6.44; 3.61 & 10.6; 0.96 \\ 
& 0. & 519.14; 519.00 & 9.44; 4.65 & 29.0; 3.67
\end{tabular}
\end{center}
\end{table}
\newpage 
\begin{table}[tb]
\caption{Same as Table I but for a SIMOX sample with 50 nm top-layer
thickness. The splitting of triplet was not resolved }
\label{tb2}
\begin{center}
\begin{tabular}{|c|c|c|c|}
Distance from & Lines $\Gamma_{25}^{\prime}$ & Widths & Interactions \\ 
interface, $\mu$m & $\omega_{x}$ & $\Gamma_{x}$ & $B_{xx}$ \\ \hline
20.0 & 520.0 & 3.40 & 0 \\ 
5.0 & 520.01 & 3.50 & 0.20 \\ 
2.0 & 520.02 & 3.60 & 0.38 \\ 
1. & 519.76 & 3.70 & 0.58 \\ 
0. & 519.50 & 3.79 & 0.77 \\ 
0. & 519.30 & 3.79 & 0.77
\end{tabular}
\end{center}
\end{table}

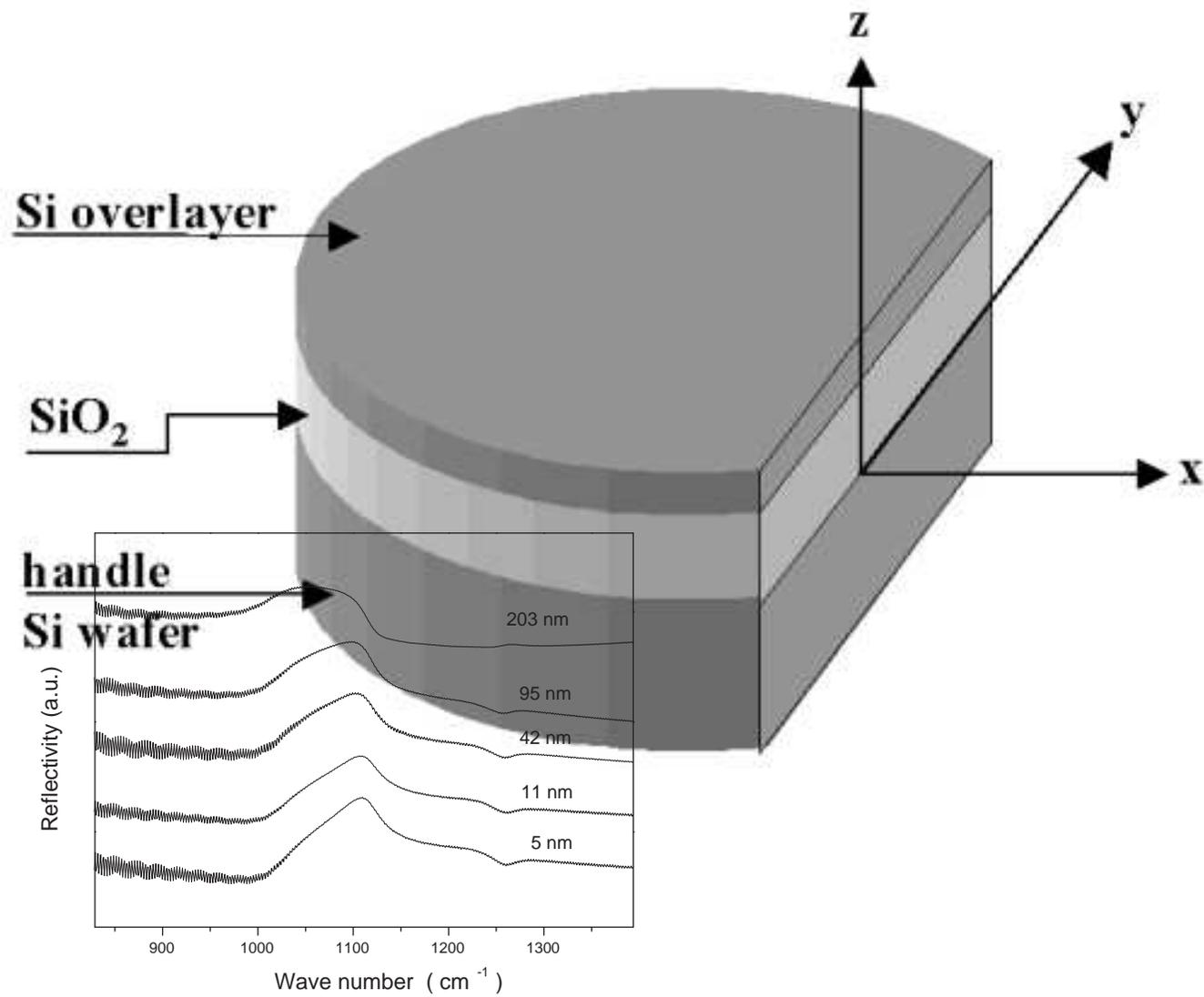
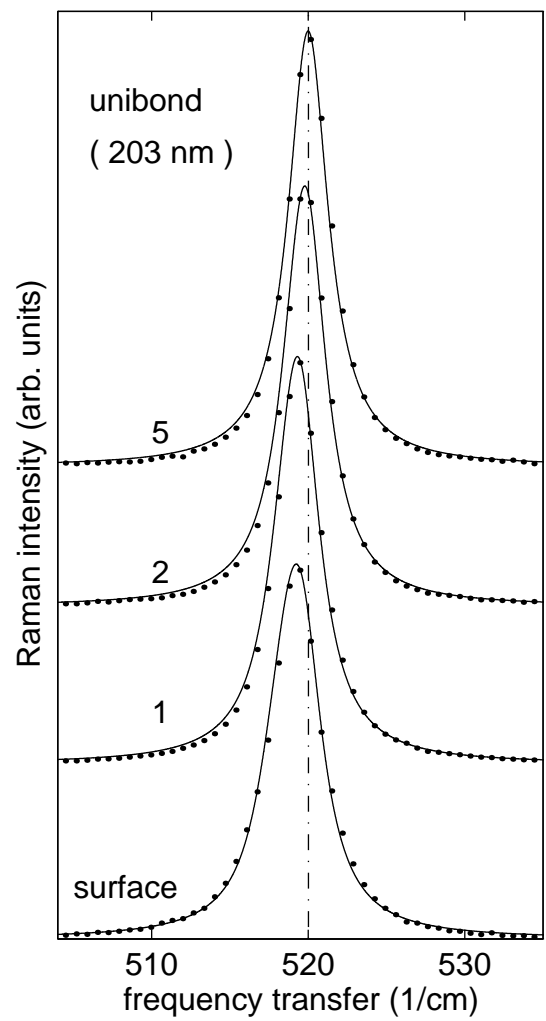
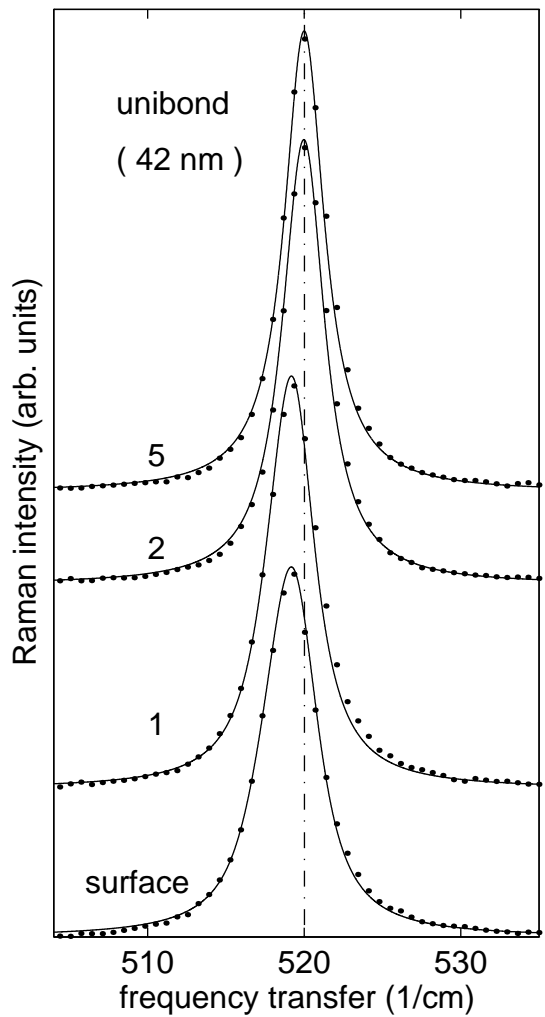
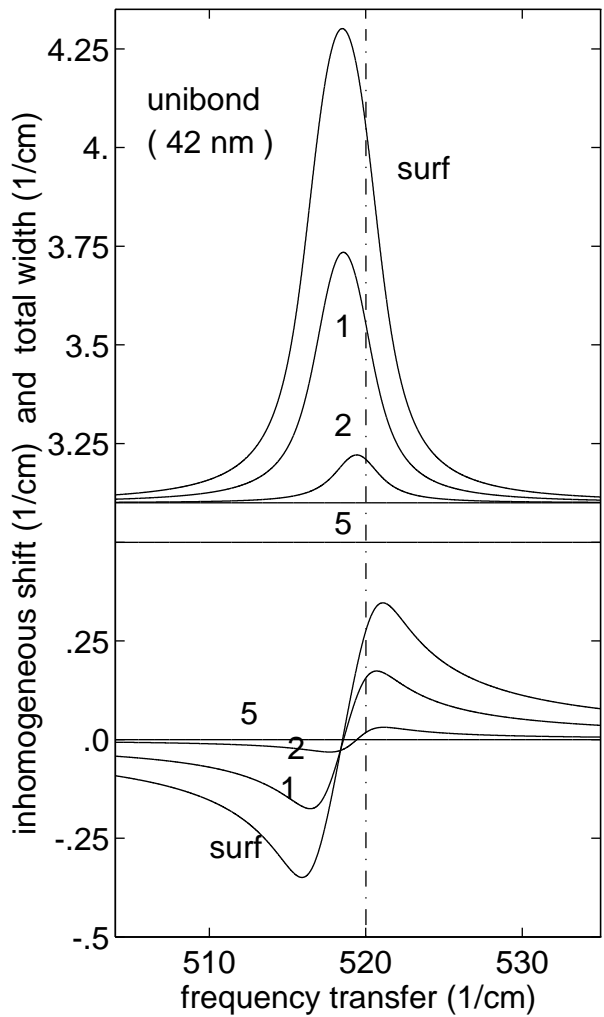
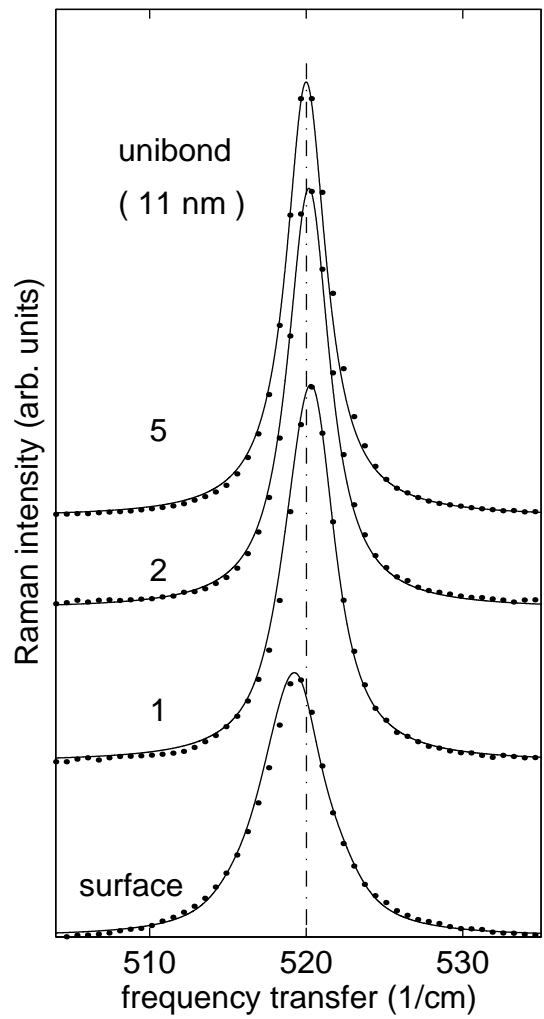
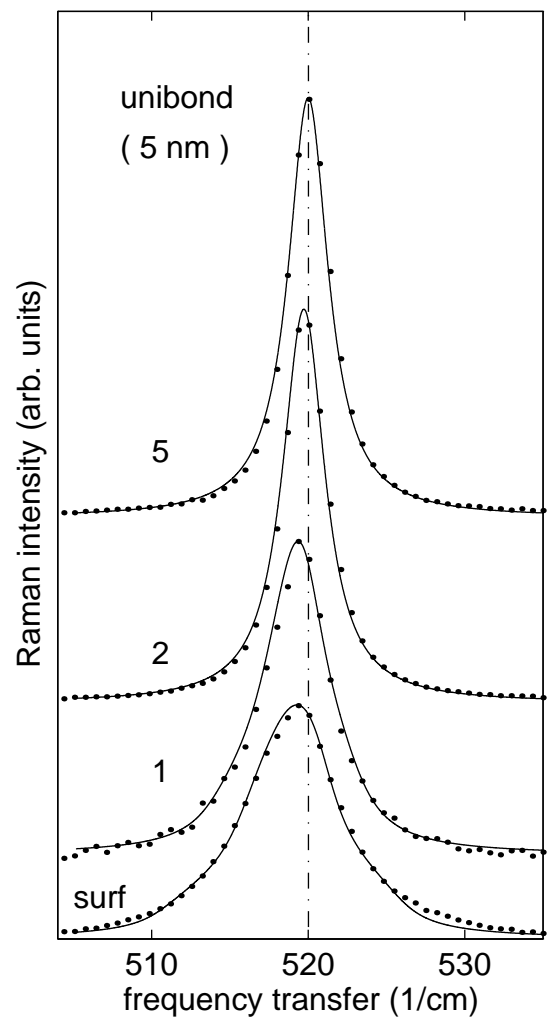
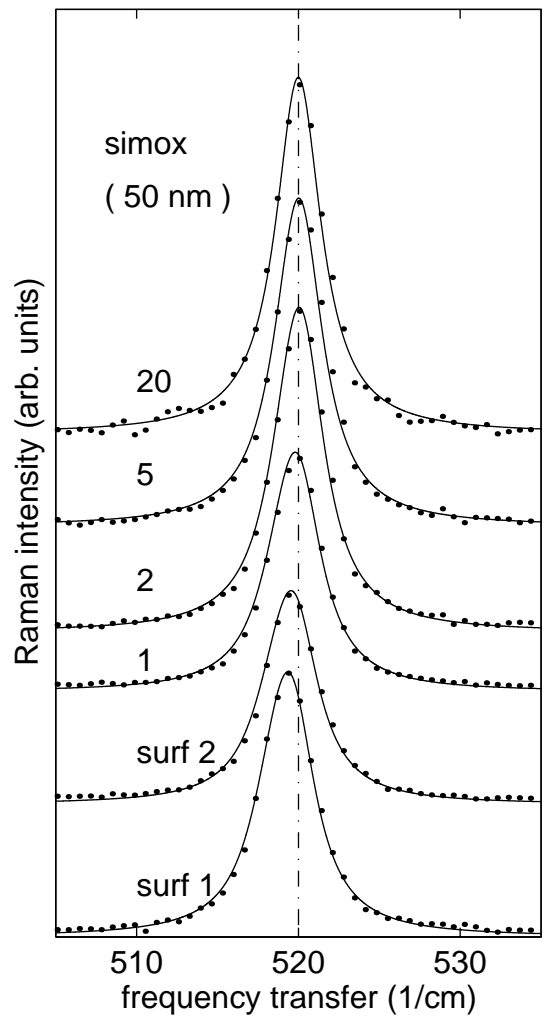
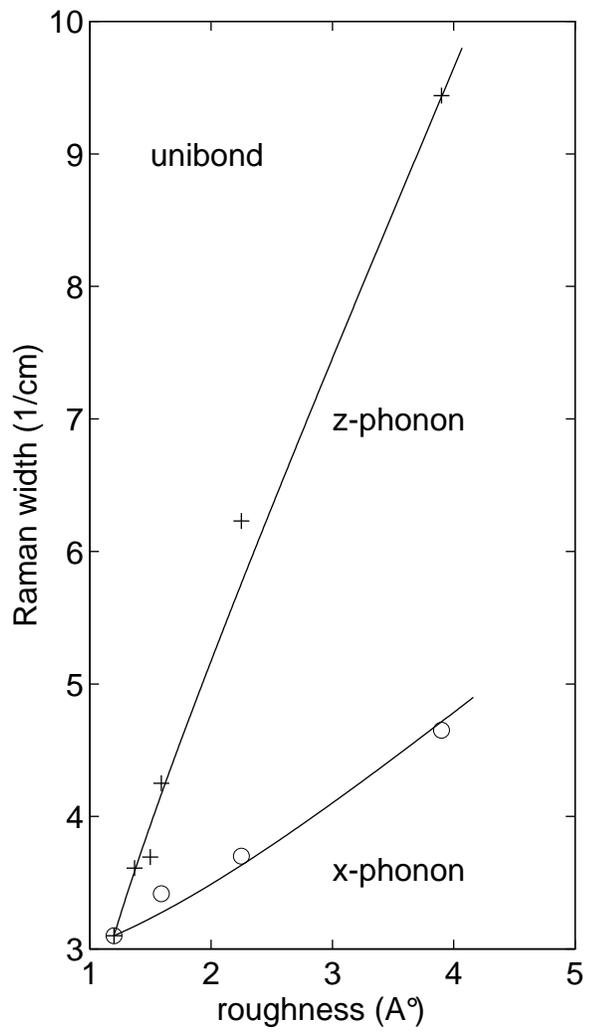
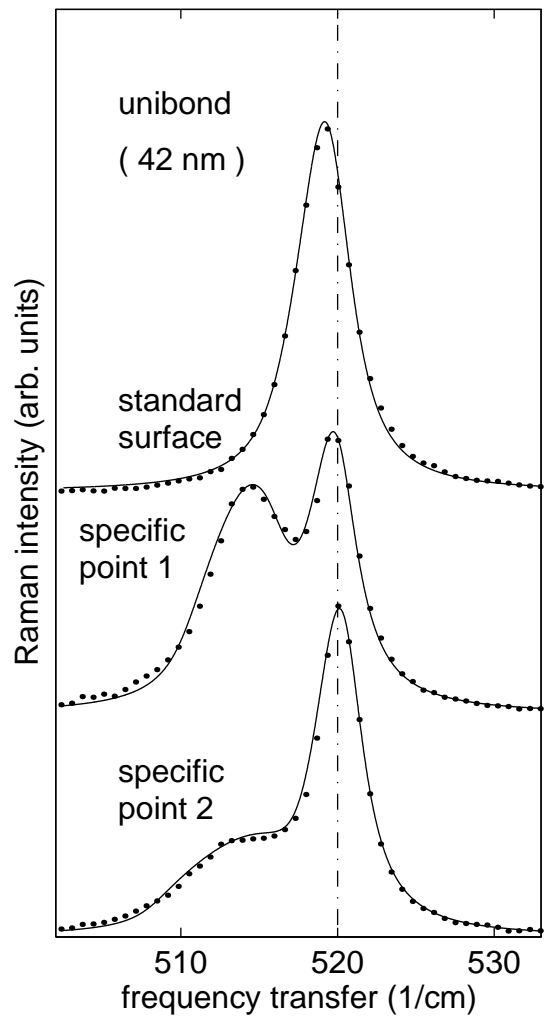
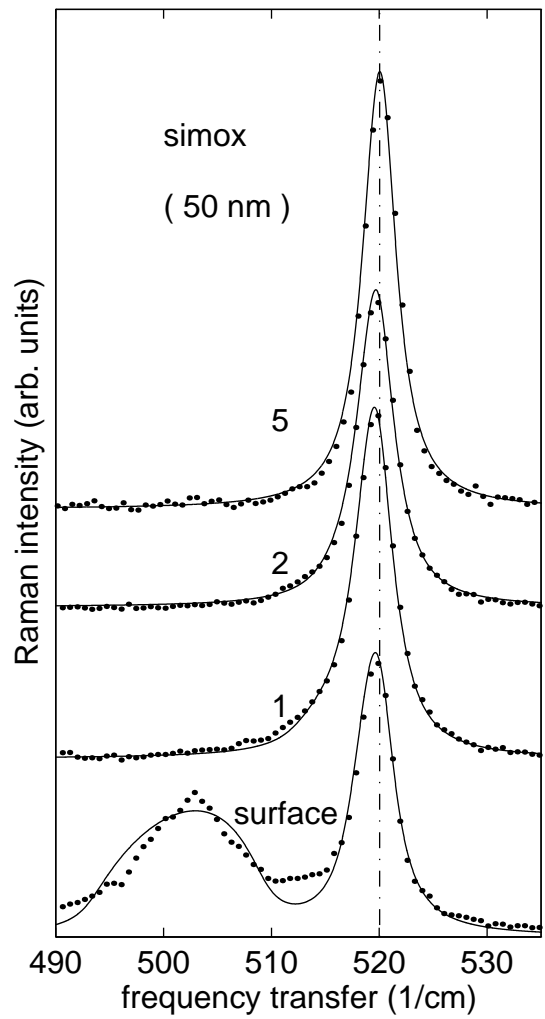
\end{document}